\documentclass[twocolumn,superscriptaddress]{revtex4}

\usepackage{graphicx,amsmath,amssymb}

\begin{document}

\title{Observing the intrinsic linewidth of a quantum-cascade laser: beyond the Schawlow-Townes limit.}

\author{S.~Bartalini}\email[e-mail: ]{saverio.bartalini@ino.it}
\author{S.~Borri}
\author{P.~Cancio}
\affiliation{Istituto Nazionale di Ottica (INO) - CNR, Largo Fermi 6, 50125 Firenze FI, Italy}
\affiliation{European Laboratory for Non-linear Spectroscopy (LENS), Via Carrara 1, 50019 Sesto Fiorentino FI, Italy}

\author{A.~Castrillo}
\affiliation{Dipartimento di Scienze Ambientali - Seconda Universit\`a di Napoli, Via Vivaldi 43, 81100 Caserta CE, Italy}

\author{I.~Galli}
\author{G.~Giusfredi}
\author{D.~Mazzotti}
\affiliation{Istituto Nazionale di Ottica (INO) - CNR, Largo Fermi 6, 50125 Firenze FI, Italy}
\affiliation{European Laboratory for Non-linear Spectroscopy (LENS), Via Carrara 1, 50019 Sesto Fiorentino FI, Italy}

\author{L.~Gianfrani}
\affiliation{Dipartimento di Scienze Ambientali - Seconda Universit\`a di Napoli, Via Vivaldi 43, 81100 Caserta CE, Italy}

\author{P.~De~Natale}
\affiliation{Istituto Nazionale di Ottica (INO) - CNR, Largo Fermi 6, 50125 Firenze FI, Italy}
\affiliation{European Laboratory for Non-linear Spectroscopy (LENS), Via Carrara 1, 50019 Sesto Fiorentino FI, Italy}

\date{\today}

\begin{abstract}
A comprehensive investigation of the frequency-noise spectral density of a free-running mid-infrared quantum-cascade laser is presented for the first time. It provides direct evidence of the leveling of this noise down to a white noise plateau, corresponding to an intrinsic linewidth of a few hundred Hz. The experiment is in agreement with the most recent theory on the fundamental mechanism of line broadening in quantum-cascade lasers, which provides a new insight into the Schawlow-Townes formula and predicts a narrowing beyond the limit set by the radiative lifetime of the upper level.
\end{abstract}

\maketitle

Quantum-cascade lasers (QCLs) are nowadays emerging as leading sources in the infrared (IR) range for a huge variety of applications. We are witnessing an extraordinary technological boost of such devices (compact units with room-temperature operation, external-cavity designs, lasers arrays, etc.) that, till now,  has not been focused to improving their intrinsic features. 
In contrast, a thorough knowledge and control of QCLs frequency fluctuations can enable their application in demanding fundamental research fields, as well as in applications based on high-sensitivity trace-gas detection.

The unique features of QCLs mark significant differences with respect to common diode lasers, also concerning the line-broadening mechanisms. Progress towards a deeper understanding of these differences are leading to a reformulation, for QCLs, of the Schawlow-Townes (S--T) formula \cite{Schawlow58}, in analogy with what happened for semiconductor (SC) lasers. Indeed, since the advent of diode lasers in 1962, twenty years of theoretical and experimental activity \cite{Hall62,Lax67,Lang73
} were needed to formalize the theory of their linewidth \cite{Henry82}. The S--T equation was corrected by the so-called Henry \textit{linewidth enhancement factor} $\alpha_e$:
\begin{equation}
\delta \nu = \frac{v_g^2 \, h \nu \, \alpha \,  \alpha_m \, n_{sp}}{4 \pi \, P_o}(1+\alpha_e^2),
\label{eq:Henry}
\end{equation}
where $v_g$,  $n_{sp}$, $\alpha$, $\alpha_m$ and $P_o$ are the group velocity, the spontaneous emission factor (or population inversion parameter), the total cavity losses, the mirror losses and the output power, respectively.  The total cavity losses $\alpha$ include the mirror losses, the waveguide losses and the absorption losses in the highly doped injector.
The $\alpha_e$ factor accounts for the effect of refractive index variations caused by electron density fluctuations, and finally explains the line broadening shown by SC lasers. 
For QCLs, on the contrary, $\alpha_e$ close to zero was predicted, since their refractive index variations are negligible at the peak of the gain spectrum \cite{Faist94}. Several experiments confirmed the small values of $\alpha_e$ \cite{VonStaden06, Aellen06, Kumazaki08}, supporting the widespread opinion that the intrinsic linewidth of QCLs should be closely described by the standard S--T formula. 
\begin{figure}[hb]
	\begin{center}
			\includegraphics[width=.7 \columnwidth]{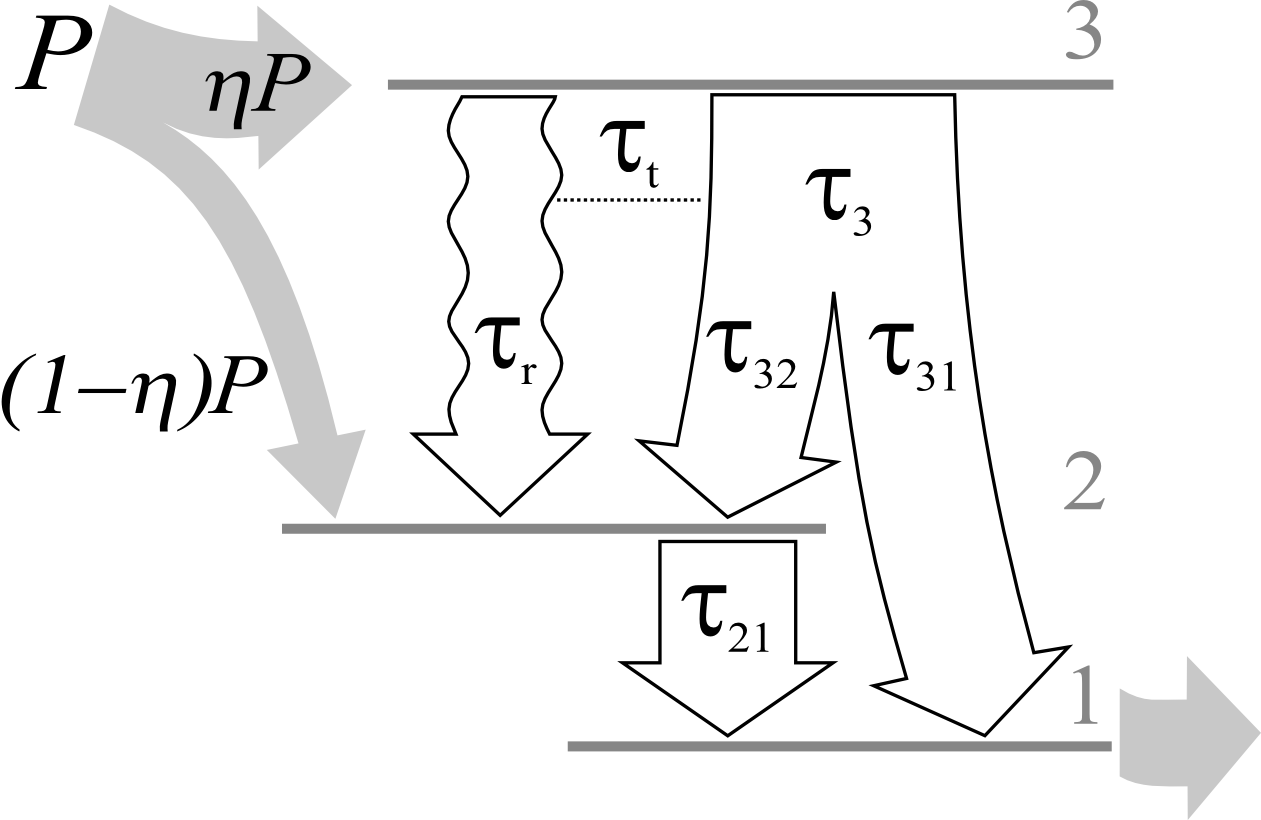}
	\end{center}	
	\caption{Three-level model of one stage of the QCL active region. The lasing transition is between levels 3 and 2, with a radiative lifetime $\tau_r$, while $\tau_3 = (1/\tau_{31}+1/\tau_{32})^{-1}$ is the lifetime of the non-radiative relaxation processes of the upper level. Its total lifetime is $\tau_t =(1/ \tau_r +1/ \tau_3)^{-1}$. $P=I/e$ is the pump rate corresponding to the current $I$, and $\eta$ is the injection efficiency from the previous stage.}
	\label{fig:livelli}
\end{figure}

Indeed, only a very recent theoretical work by Yamanishi and coworkers \cite{Yamanishi08}, based on rate equation analysis of a three-level model of a QCL (see Fig. \ref{fig:livelli}), provided a full explanation of the physics that governs the linewidth of QCLs. This work carried out a reformulation of the S--T equation in terms of the characteristic parameters of the QCL medium, leading to a remarkable prediction: in QCLs the linewidth should overcome the limit set by the S--T formula for conventional bipolar SC lasers, giving much narrower emission features. This work introduces the novel concept of ``effective coupling'' ($\beta_{eff}$) of the spontaneous emission, given by the ratio of the spontaneous emission rate coupled into the lasing mode ($\beta/\tau_r$) to the total relaxation rate ($1/\tau_t$) of the upper level. 
The modified formula, tailored to a QCL typical design, is \cite{Yamanishi08}:
\begin{equation}
		 \delta \nu = \frac{1}{4 \pi}\frac{\gamma \, \beta_{eff}}{\left(1-\epsilon \right)} 
		 \cdot \left[ \frac{1}{\left(I_o/I_{th}-1\right)} + \epsilon \right] \cdot (1+\alpha_e^2) ,
	\label{eq:eq16a}
\end{equation}
with:
\begin{equation}
		\epsilon = \frac{\tau_{21}\tau_{31}}{\eta \tau_t (\tau_{21}+\tau_{31})}.
	\label{eq:epsilon}
\end{equation}
The inverse photon lifetime $\gamma = v_g \alpha$, the upper level injection efficiency $\eta$,  the lifetimes $\tau$ of the involved energy levels and $\beta_{eff}$ itself are fixed by design, while the only experimental parameter is the ratio of the operating current $I_o$ to the threshold current $I_{th}$. Using the $\tau$ values provided by Ref. \cite{Hofstetter01} for a QCL similar to that used in our experiment, the expected linewidth can be derived for each $I_o/I_{th}$ value.  
According to Ref. \cite{Yamanishi08}, we calculate $\gamma \simeq 1.3 \cdot 10^{11}$~s$^{-1}$,  $\tau_r \simeq 7.5$~ns and $\beta_{eff} \simeq 1.6 \cdot 10^{-8}$, while for the injection efficiency $\eta \simeq 0.7$ we base on standard values \cite{Faist94,Capasso02}. It is worth noting that our QCL is based on a two-phonon design, that is better described by a four-level scheme \cite{Faist02}. However, the difference with the three-level scheme involves only the bottom level 1, which splits into two levels, and essentially leads to a more efficient depopulation of level 2. This is taken into account by using effective $\tau$ parameters in the three-level simplified scheme presented above.

Unlike SC lasers, in QCLs the non-radiative decay lifetime of the upper level 3 is many orders of magnitude smaller than the spontaneous emission lifetime (in our case $\tau_3 \simeq 1.3$~ps $\ll \tau_r \simeq 7.5$~ns). This is also proven by the very weak electroluminescence shown by below-threshold QCLs. Above threshold, the fast non-radiative relaxation processes run in parallel with the spontaneous emission, and efficiently compete with it. As a result the noise associated with spontaneous emission is strongly suppressed, leading to a major linewidth reduction.
Due to this effect the dependence of the linewidth on the operating current, in the new model, follows the same hyperbolic law of the S--T equation, but with a sharper behavior. This means that the narrow-linewidth regime is achieved just above threshold.

To date, however, no analysis of the frequency-noise power spectral density (PSD) of QCLs has given a direct evidence of the theoretical predictions. This was probably due to the challenging set-up required to measure such a tiny and elusive intrinsic linewidth. On the other hand, some complementary results are supporting this hypothesis, such as the achievement of sub-kHz linewidths in frequency-stabilized QCLs \cite{Williams99,Taubman04}. Nonetheless, the mechanisms at the basis of QCLs frequency-noise figures still need to be investigated, as well as their correlation with external noise sources, such as the driving current noise. 
Our study provides, for the first time, a complete overview on the frequency-noise PSD of a free-running mid-IR QCL, giving a quantitative evaluation of an intrinsic linewidth that is shown to be consistent with the theoretical prediction.

The frequency-noise PSD provides, for each frequency, the amount of noise contributing to the spectral width of the laser emission. Its determination gives the most direct way to characterize  the spectral purity, since it allows to derive the laser emission spectrum over  any accessible timescale \cite{Elliott82}. It also allows to calculate the linewidth reduction achievable by a frequency-locking loop, once its bandwidth is known. Finally, it enables the tracing of spurious noise sources that can be removed in order to measure only the intrinsic laser noise. 

In our experiment, intensity measurements are performed to retrieve information in the frequency domain by converting the laser frequency fluctuations into detectable intensity variations. 
Given the low-noise nature of the measurement, the converter must introduce a negligible noise providing, at the same time,  a gain factor suitable for a good detection. To this purpose, the side of a molecular transition has proven to be the most suitable tool: the spectrum of the intensity transmitted by such discriminator is acquired, with the laser frequency ($\nu_0$) stabilized at the half height position, and reproduces the spectrum of the laser frequency fluctuations, ``amplified" by the slope of the absorption profile.
In combination with fast detection and low-noise fast-Fourier-transform (FFT) acquisition, this technique enables spectral measurements spanning over 7 frequency decades (from 10~Hz to 100~MHz) and 10 amplitude decades.

The QCL source is of the distributed feed-back (DFB) type ($\lambda = 4.33$~$\mu$m) with single-mode continuous-wave operation at temperatures in the range between 81 and 92~K. Mounted inside a liquid-nitrogen cryostat, it is driven by a home-made ultra-low-noise current supply.
Similarly to what already described elsewhere \cite{Bartalini07}, direct-absorption spectroscopy of CO$_2$ is performed in a 10-cm-long cell, and the signal is acquired by a HgCdTe photovoltaic detector with a nominal bandwidth of 200~MHz and an output voltage noise of 3$\cdot$10$^{-8}$~V/$\sqrt{\textrm{Hz}}$.
The molecular ro-vibrational transition working as frequency-to-amplitude converter is the $(01^11-01^10)$ P(31), at a constant pressure of 1~mbar. 
Both the threshold current $I_{th}$ and the operating current $I_o$ corresponding to the optical frequency $\nu_0$ depend on temperature, as well as the laser emission power. 
The final detection sensitivity, in frequency-noise PSD units, is about 10~Hz$^2$/Hz. 

\begin{figure}[htb]
	\begin{center}
			\includegraphics[width=\columnwidth]{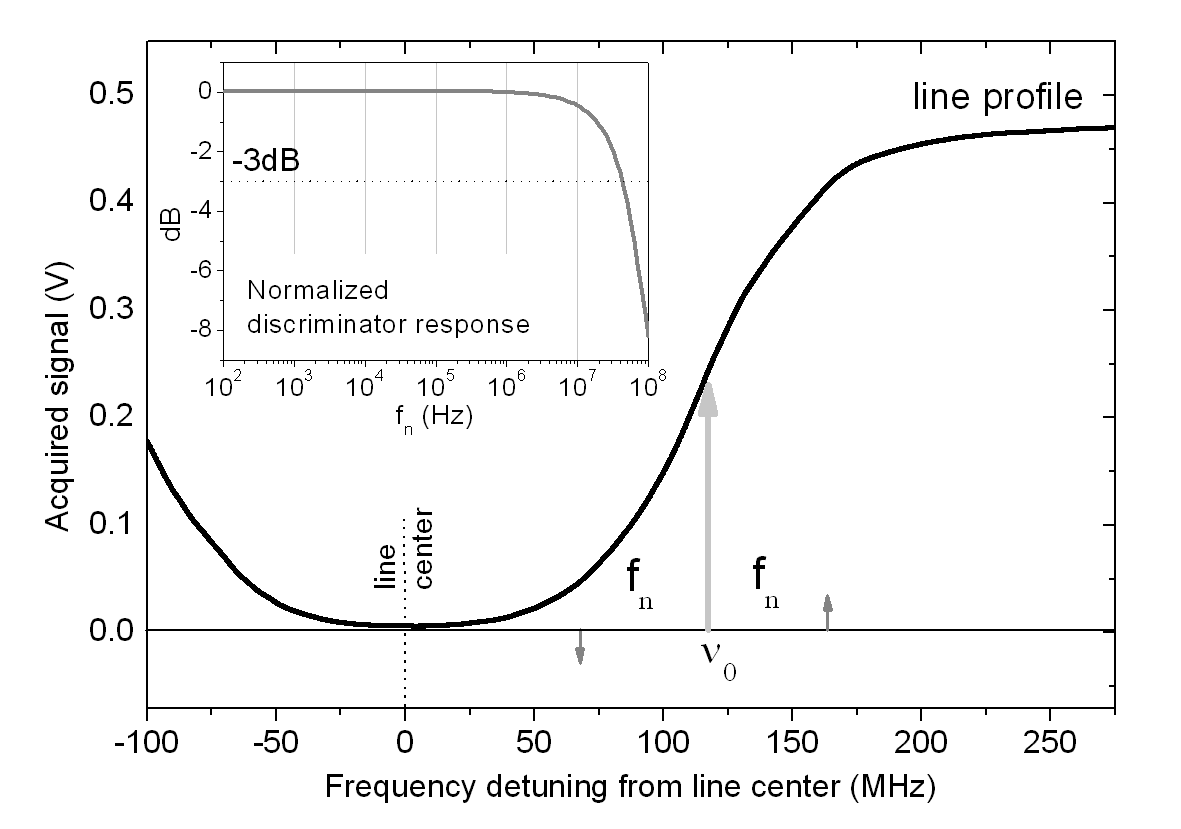}
	\end{center}	
	\caption{Principle of the frequency-noise PSD measurement. The black line is the molecular absorption profile, the light-grey arrow is the laser frequency $\nu_0$ and the small dark-grey arrows represent the noise component at frequency $f_n$. The strong-absorbance regime allows to maximize both the slope (i.e. the gain) and the dynamic range of the measurement. \textit{Inset:} Spectral response of the frequency-to-amplitude converter. The flat response region corresponds to the region where the absorption profile can be approximated as linear. The 3-dB cut-off lies between 40 and 50~MHz.}
	\label{fig:riga}
\end{figure}

In order to correctly understand the spectrum up to frequencies where the discriminator response is not linear, a study of its transfer function is required (see Fig.~\ref{fig:riga}). Any spectral component of the laser frequency noise can be treated as a sideband associated to a frequency-modulated laser beam, so that the formalism of frequency-modulation spectroscopy can be used \cite{Silver92}. In this way a cut-off at about 50~MHz is obtained (see inset of Fig. \ref{fig:riga}), so that in the measurement range (100~MHz) this effect is taken into account for reconstructing the correct profile of the frequency-noise PSD. Above 100 MHz the sensitivity limit is approached due to the discriminator cut-off. 

\begin{figure}[htb]
	\begin{center}
		\includegraphics[width=\columnwidth]{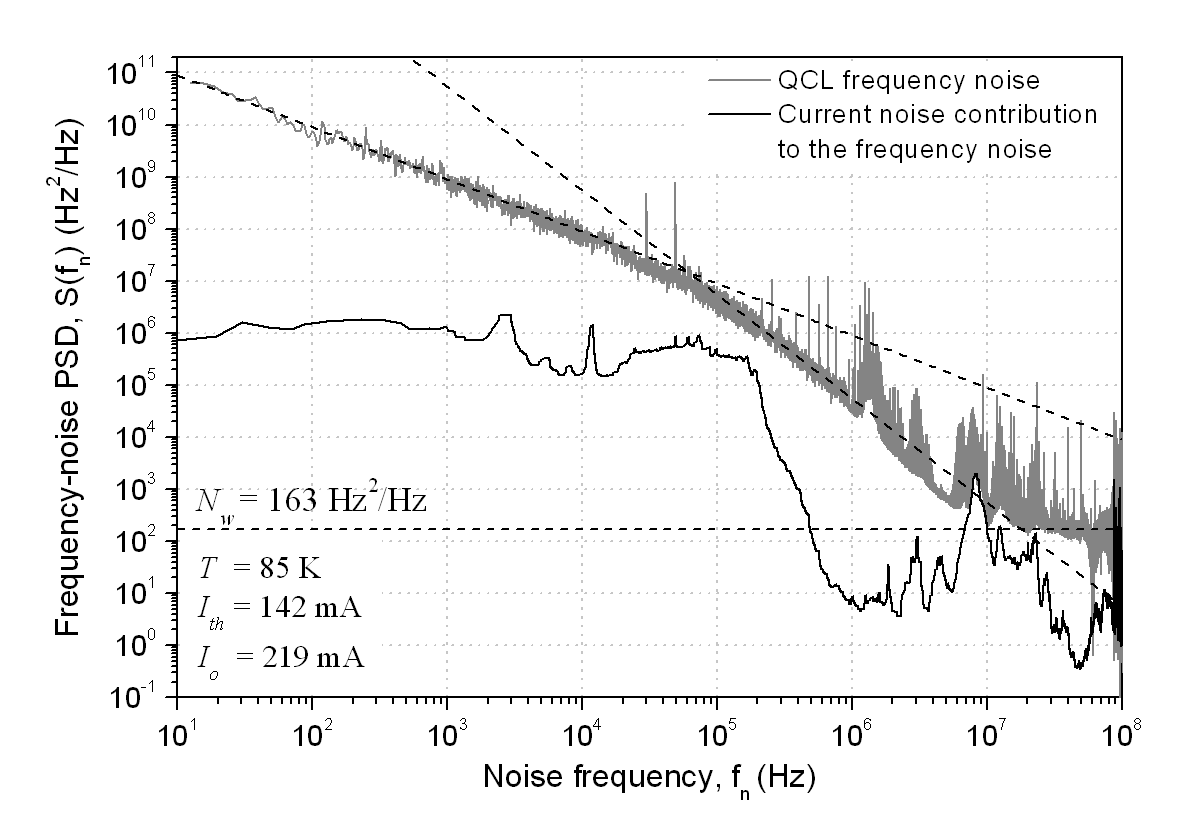}
	\end{center}
	\caption{Frequency-noise PSD of the QCL (grey) acquired between 10~Hz and 100 MHz, compared with the expected contribution arising from the driving-current noise (straight black). 
The dashed lines show the three main trends: the 1/f and 1/f$^2$ trends fit well the experimental data up to a few MHz, while at higher frequencies the flattening to a white-noise level $N_w$ is evident.}
	\label{fig:noise}
\end{figure}

The full spectrum of the QCL frequency-noise PSD shown in Fig.~\ref{fig:noise} is obtained by sticking together several acquisitions on smaller spectral windows, in order to ensure a high overall resolution. Although residual external noise gives rise to the sharp peaks visible throughout the trace, it is possible to clearly recognize three distinct domains, defined by different noise dependences on frequency:

10 Hz $\div$ 100 kHz:	$1/f$  trend

100 kHz $\div$ 10 MHz: 	$1/f^2$ trend

10 $\div$ 100 MHz: 		asymptotic flattening.

A first comment concerns the contribution from the supply current to the frequency noise. It has been estimated by measuring the current-noise PSD of the driver over the entire frequency range. As shown by the black trace in the graph, its contribution is negligible, so that the measured spectrum can only be attributed to the laser.

The trend of the frequency noise PSD, here observed for the first time in the whole significant range, can shed light on fundamental heat or charge transport mechanisms occurring in the laser medium. In particular, in the low-frequency region, the $1/f$ behavior and the slope change to $1/f^2$ at around 100 kHz still need further investigation.

Above 10 MHz, the noise profile deviates significantly from the $1/f^2$ trend,  and its bending suggests a flattening to a white-noise level ($N_w$). 
According to frequency-noise  theory, the laser power spectrum corresponding to the white component $N_w$ of the frequency noise is purely Lorentzian, with a full width at half maximum (FWHM) $\delta \nu = \pi N_w$ \cite{Zhu93}. 
The measurement shown in Fig. \ref{fig:noise}, taken for $I_o/I_{th} = 1.54$, yields an intrinsic linewidth $\delta \nu~=~510 \pm 160$~Hz,
that is in good agreement with the linewidth  $\delta \nu_Y=468$~Hz predicted by Yamanishi's model for the same $I_o/I_{th}$.

\begin{figure}[htb]
	\begin{center}
		\includegraphics[width=\columnwidth]{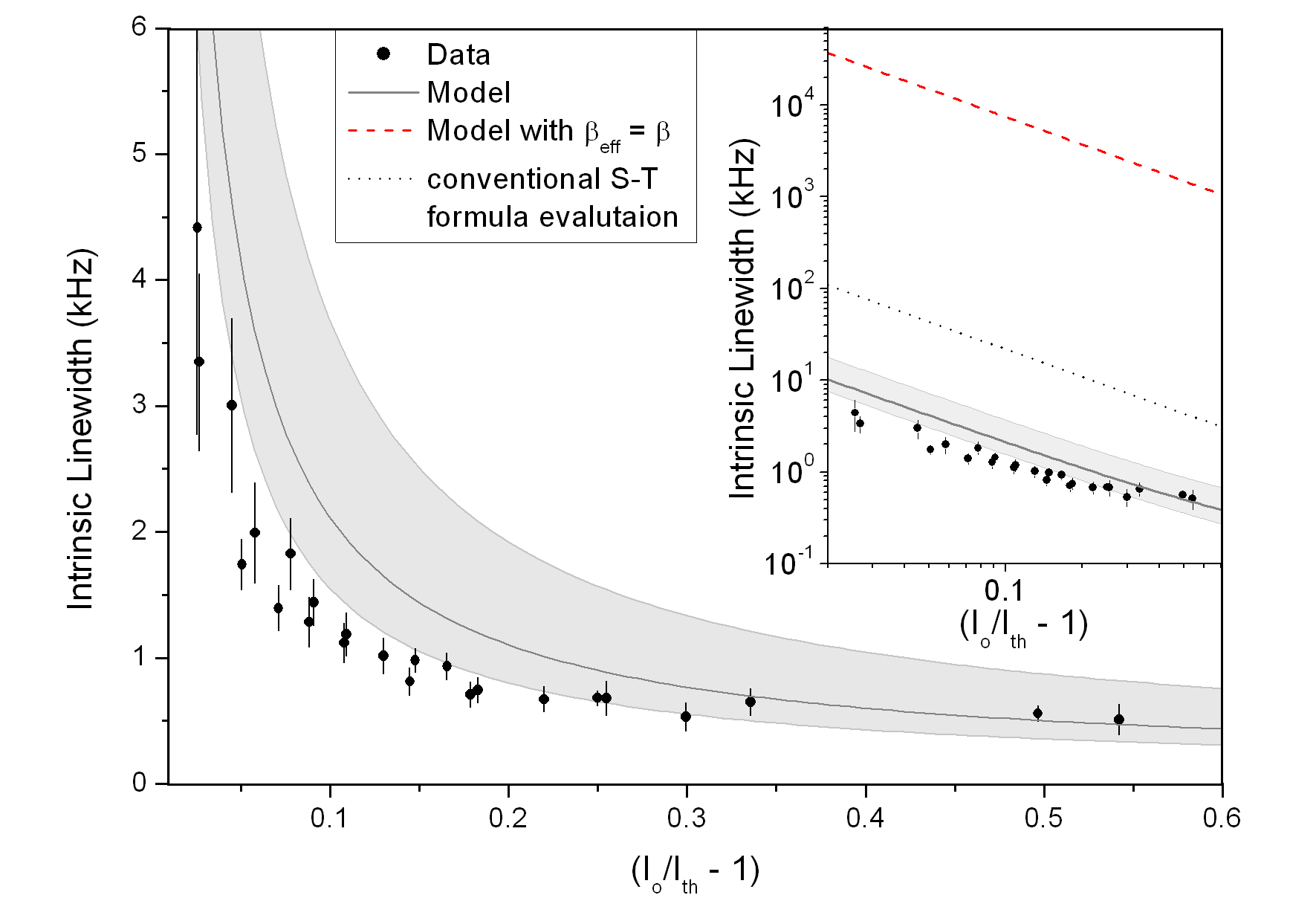}
	\end{center}
	\caption{The experimental data set (black dots) is compared with the prediction of the theoretical model by Yamanishi \textit{et al.} (Eq. \ref{eq:eq16a}).  The dark grey curve corresponds to the set of parameters considered as the most realistic for our device:  $\tau_{21}=0.25$~s, $\tau_{31}=2.0$~ps, $\tau_{32}=3.4$~ps, $\tau_t=1.3$~ps (see Ref. \cite{Hofstetter01}). The grey area in the plot covers the uncertainty space, as inferred by the parameters variability for similar devices reported in literature: $0.15$~ps~$< \tau_{21} < 0.35$~ps, $1.8$~ps~$<\tau_{31}<4.5$~ps,  $2.5$~ps~$<\tau_{32}<4.3$~ps, 7~ns~$ < \tau_r <$~8~ns, $0.6 < \eta < 0.8$ and  $1.2 \cdot 10^{11}$~s$^{-1} < \gamma <  1.4 \cdot 10^{11}$~s$^{-1}$.  \textit{Inset:} comparison with the model calculated for $\beta_{eff} = \beta$ and with the conventional S--T formula evaluation described in the text.
	}
	\label{fig:noisetrend}
\end{figure}

It is worth making a comparison with the linewidth predicted by the same model for $\beta_{eff}=\beta$. This is equivalent to switching off the competitive non-radiative relaxation channel, obtaining $\tau_t=\tau_r$. The linewidth results in the MHz range, the typical value for conventional bipolar SC lasers (see inset of Fig.~\ref{fig:noisetrend}).
It is also interesting  to consider a conventional estimation of the standard S\nobreakdash--T formula (Eq. \ref{eq:Henry}), based on a rough evaluation of the $\alpha_m$ parameter as the bare Fabry-Perot facets loss. The value of 4~kHz, obtained for our measured output power $P_o \simeq 6$~mW, is consistent with the evaluated linewidths commonly found in literature \cite{Williams99,Myers02}. The overestimation is mainly caused by the difficulty in retrieving the correct value of the laser internal power for a DFB QCL. The Yamanishi's model overcomes this problem providing an explicit formula entirely based on the laser fabrication parameters. 

For an exhaustive comparison with the new model, the frequency noise has been measured, at different operating currents, in the region where the frequency-noise PSD is flat ($f>30$~MHz). By changing the temperature, the ratio $I_o/I_{th}$ changes consequently,
allowing to measure the dependence of the intrinsic linewidth on $I_o/I_{th}$.
In Fig.~\ref{fig:noisetrend} the comparison between measurement and theory is shown. The qualitative behavior is well reproduced, as evidenced by the presence of the divergence for close-to-threshold currents and by its hyperbolic trend. 

In addition, experimental linewidth values are systematically smaller than predictions, that is consistent with $\alpha_e$ close to zero.
Further improvements of the measurement sensitivity, in particular by using faster detectors with larger detectivities (such as quantum well infrared photodetectors), will enable a more accurate investigation of the intrinsic linewidth and might help in developing a better model. Intriguing perspectives could also be opened by performing similar measurements on THz QCLs, as well as new predicted effects, such as line-broadening by black body radiation \cite{Yamanishi08}, are still awaiting confirmation.

In conclusion, our work provides a thorough overview of the frequency-noise spectral density of a quantum-cascade laser up to 100 MHz, confirming that QCLs have unique noise features. We also provide experimental evidence of a linewidth narrowing beyond the limit set by the spontaneous emission rate, which was thought to be a fundamental limit for all lasers. 
The measured intrinsic linewidths are comparable with the natural linewidth of molecular IR ro-vibrational transitions. Therefore, similarly to what  visible/near-IR-emitting diode lasers have represented in the last 20 years for the progress of atomic physics, mid-IR QCLs can become unique tools for investigating and harnessing molecules with unprecedented precision levels.


We gratefully acknowledge M. Yamanishi from Hamamatsu Photonics for his invaluable support on the theory, F. Marin from University of Firenze for fruitful discussions, and  M. Giuntini and A. Montori from LENS for their help on driving electronics. We thank Alpes Laser for kindly providing additional information on the laser.
This work was partially supported by Ente Cassa di Risparmio di Firenze, by Centro Regionale di Competenza INNOVA, and by CNR and the
other European research funding Agencies participating to the European Science Foundation EUROCORES Program EUROQUAM-CIGMA.

\end{document}